# National research assessment exercises: the effects of changing the rules of the game during the game[1]

*Giovanni Abramo*[a,b,*], *Ciriaco Andrea D'Angelo*[b] *and Flavia Di Costa*[b]

[a] National Research Council of Italy

[b] Laboratory for Studies of Research and Technology Transfer
School of Engineering, Dept of Management
University of Rome "Tor Vergata"

**Abstract**

National research evaluation exercises provide a comparative measure of research performance of the nation's institutions, and as such represent a tool for stimulating research productivity, particularly if the results are used to inform selective funding by government. While a school of thought welcomes frequent changes in evaluation criteria in order to prevent the subjects evaluated from adopting opportunistic behaviors, it is evident that the "rules of the game" should above all be functional towards policy objectives, and therefore be known with adequate forewarning prior to the evaluation period. Otherwise, the risk is that policy-makers will find themselves faced by a dilemma: should they reward universities that responded best to the criteria in effect at the outset of the observation period or those that result as best according to rules that emerged during or after the observation period? This study verifies if and to what extent some universities are penalized instead of rewarded for good behavior, in pursuit of the objectives of the "known" rules of the game, by comparing the research performances of Italian universities for the period of the nation's next evaluation exercise (2004-2008): first as measured according to criteria available at the outset of the period and next according to those announced at the end of the period.

# 1. Introduction

The use of national evaluation exercises is becoming ever more common and frequent. These exercises furnish a comparative measure of research performance for the research institutions of a given nation. As such, they represent a stimulus instrument towards improved productivity, which becomes still more useful when the results are used to inform selective funding by government (Debackere and Glänzel, 2004; Rousseau and Smeyers, 2000). They also serve to reduce information asymmetry between the supply of new knowledge and the demand from students, companies, and others. Research assessment exercises are essentially policy instruments, in which the government can select evaluation criteria to influence and direct the strategies and actions of research institutions. For example, evaluating the research products of a limited number of researchers per university can support goals of reinforcing centers of excellence. Vice versa, evaluating all the researchers of a university supports goals of raising the performance level of all research staff.

The response by university administrators to specific evaluation criteria consists of formulating and applying internal incentive systems, which take some time before they can give the desired results. Because of this, one of the most common complaints from the subjects evaluated concerns the frequent changes in criteria. The rules of the game could be changed for several motives: i) based on accumulated experience, both implementation and methodology of the evaluation can be amended or refined ii) advancements in knowledge and techniques in the broader field of research evaluation can offer new indicators and measurement methods; iii) resources available for implementation of the exercises can vary, and influence planning; iv) policy aims and objectives can vary, along with associated evaluation criteria.

Studies have examined both methodology and applications for the analysis of university research productivity and impact (Abramo and D'Angelo, 2011; Gómez et al., 2009; Kao and Pao, 2009; Abramo et al., 2008), but also questions about possible effects induced by incentive schemes on researcher behavior (Moed, 2008; Gläser, 2007; Laudel, 2006; Butler, 2003). Performance-based research funding can lead to changes in managerial practice that in turn influence researchers and the work they produce. For example, management practice can pressure researchers to conform to the directions set by the evaluation process; can link internal promotion to evaluation results; assign resources for research in a more concentrated or more diversified manner; or introduce internal evaluation-based funding (Gläser, 2007). Geuna and Martin (2003) and Liefner (2003) have conducted international-level comparative analyses of adoption of performance-based funding. At national levels, studies have been conducted in Australia and the UK to verify whether researcher behavior tends to modify towards alignment with evaluation criteria. Butler (2003) studied the effects on Australian researchers' performance following introduction of an incentive system based on publication counts as a key criterion to be used in allocation of a significant share of funds. The effects, evaluated *a posteriori*, show an increase in publication productivity between 1990 and 1998, but with a corresponding drop in relative quality at the international level.

Moed (2008) examined, for the various UK research assessment exercises (RAE) that followed one another over the 1985-2004 period, whether behavior of research staff tended to align over time with the guidelines of the evaluation exercises. When, in 1992, the RAE placed emphasis on the quantitative aspect of scientific production the response was an increase in publication numbers. However in 1996, when the focus shifted from "output counts" to "quality", there was a greater propensity to publish in journals with a higher impact factor.

While somebody welcomes frequent changes in the evaluation criteria in order to deter the subjects evaluated from adopting opportunistic behaviors, it is certainly true, and more fundamental, that the rules of the game must serve in the pursuit of policy objectives, meaning that the design of an evaluation exercise should provide criteria that are aligned and coherent with the objectives. This means that it should not occur, as is so often the case, that the evaluation criteria for research institutions are set and communicated during or at the end of the period subject to evaluation. We take the example of the Research Excellence Framework (REF), the new system for assessing the quality of research in UK higher education system, which will replace the former RAE. During 2009 the REF steering committee conducted two broad consultations on proposals for the new framework. In March 2010 there was an announcement of some resulting decisions and the next steps anticipated for designing and implementing the framework. Detailed guidance on submissions and assessment criteria will be published during 2011. Institutions will be invited to make submissions during 2013 and the assessment will take place during 2014, but will cover research activity over the five years from 2008 to 2012.

When adjustments in evaluation criteria are not communicated far in advance of the evaluation period, a clear paradox arises. Should reward go to research organizations that responded best to the criteria of the preceding exercise or to those that result as best under the new rules that emerged during or after the period of observation? In other words, should reward be for "obedience" or "disobedience"? In the Italian case, for example, the first national evaluation exercise (VTR, 2006) produced performance rankings of universities based on an evaluation of a share of their 2001-2003 product that was equal in number to 25% of each university's research staff complement in each of the 18 disciplines considered. The next evaluation exercise (VQR) will evaluate two research outputs from each scientist, for the period 2004-2008. The evaluation criteria for the first VTR clearly directed research institutions to concentrate their resources on top scientists, while the new VQR will offer reward on the basis of average performance of their research staff. Considering that the new rules were only communicated in 2010 and that awarding of selective funding will depend on the results, the paradox is evident, as are the complaints from the subjects concerned.

This work proposes to compare the research performances of Italian universities in the 2004-2008 period as measured under the criteria of the first evaluation exercise, meaning those that were known to the universities, with the measurements that result from adoption of the criteria for the next exercise, which were only communicated in 2010. In this manner we can ascertain if and to what extent certain universities would be penalized rather than rewarded for their good behavior in pursuing the objectives understood from the rules of the game that were last communicated.

The next section of the work presents the dataset and the methodology used. Sections 3 and 4 show and comment on the results obtained from the simulations of the two scenarios, while in Section 5 we carry out a comparative evaluation of the results under the two scenarios. The work concludes with a discussion of the results obtained and their relative policy implications.

## 2. Methodology: dataset and indicators

The objective of the proposed work is to compare the research performance of Italian universities over the period 2004-2008, according to two distinct evaluation frameworks: i) based on the criteria of the first Italian evaluation exercise in Italy (VTR) and ii) based on the criteria, as currently announced, for the next national exercise (VQR). Both cases provide that the universities select a certain number of products to submit to panels for each scientific area, for evaluation according to peer review criteria, additionally informed (where appropriate) by bibliometric data. The approach used in our study will be purely bibliometric: being unable to carry out a peer-review simulation on a national scale, we will substitute the peer review judgement with bibliometric impact data. This choice is amply supported by literature: at this purpose, Abramo et al. (2009) demonstrated the existence of a significant and strong correlation between ratings produced by the Italian VTR peer review (2006) and those obtained from bibliometric indicators of quality. Similar correlations were observed in other contexts, also by Moed (2009); Pendlebury (2009); Aksnes and Taxt (2004); Oppenheim and Norris (2003); Rinia et al. (1998).

The analysis will be based on data from the Italian Observatory of Public Research (ORP)[2], in turn derived from the author listings of the Thomson Reuters Web of Science (WoS). The ORP censuses the scientific product of all Italian public research institutions, along with the relative citations observed as of 30/06/2009. Beginning from the raw data for Italy indexed in the WoS, and applying a complex algorithm for reconciliation of the author's affiliation and disambiguation of the true identity of the authors, the ORP attributes each publication (article, review, and conference proceeding) to the university scientists that produced it, with an error of less than 5% (D'Angelo et al., 2011).

For robustness reasons, we limited the analysis to the hard sciences. So, the field of observation consists of all Italian universities active in: Mathematics and computer science; Physics; Chemistry; Earth science; Biology; Medicine; Agriculture and veterinary science; Industrial and information engineering[3]. In the Italian university system such disciplines are named Universities Disciplinary Areas (UDAs) and account for more than 60% of total academic research staff. Moreover, 95% of the products presented by universities to the 2001-2003 VTR in these UDAs are indexed in the ORP.

The evaluation of individual products will be conducted by application of the Article Impact Ranking (AIR). This indicator is measured on a 0 – 100 scale, for comparison to the citation distribution of Italian publications of the same year and falling in the same WoS subject category[4]: a value of 90 indicates that only 10% of the national publications for the same year and in the same WoS subject category[5] show a higher number of citations.

The simulations will be based on a five-step procedure:

---

[2] www.orp.researchvalue.it (last accessed on 2 February, 2011).
[3] Civil engineering and architecture were not considered because the WoS listings are not sufficiently representative of research output in this area.
[4] A complete list is available on http://science.thomsonreuters.com/cgi-bin/jrnlst/jlsubcatg.cgi?PC=D. Last accessed on 2 February, 2011.
[5] For publications in multi-category journals, AIR is calculated as the weighted average of the values for each subject category, with weighting equal to the average citation intensity in each single category.

- Build universities' scientific portfolios, listing all papers authored by scientists of each university, in the five years under observation;
- Classify such portofolios according to the UDA in which the author operates[6];
- For each university and UDA, select the best publications according to article impact rankings;
- Attribute to selected publications a rating based on VTR/VQR criteria;
- Average selected publications ratings and rank universities in each UDA.

## 3. University research performance rankings based on VTR criteria

The VTR provided that, for the evaluation, each university would select a number of publications for the 2001-2003 period equal to 25% of their complement of research staff for each area. Given the communication of these evaluation criteria, which support the goal of developing excellence in universities, we assume that these institutions would subsequently have concentrated their resources on their individual points of excellence, in order to show well in the next evaluation exercise. Thus we elaborate a ranking of the universities according to the criteria of the past VTR, but for the subsequent quinquennium. In the context under study, where we intend to simulate a bibliometric exercise for a period of five years (2004-2008) rather than the three years of the VTR, we consider a share of total products equal to 50% of the research staff of each UDA: Table 1 shows the numerosity and representativeness of this share with respect to the total scientific products indexed in the ORP, for each UDA.

| UDA | No. of universities | Research staff (a) | Publications selected (b = a/2) | Total publications (c) | b/c |
|---|---|---|---|---|---|
| Mathematics and computer science | 57 | 3,288 | 1,644 | 14,038 | 11.7% |
| Physics | 59 | 2,576 | 1,288 | 22,367 | 5.8% |
| Chemistry | 58 | 3,241 | 1,621 | 24,569 | 6.6% |
| Earth sciences | 47 | 1,275 | 637 | 4,639 | 13.7% |
| Biology | 61 | 5,198 | 2,599 | 28,021 | 9.3% |
| Medicine | 53 | 11,137 | 5,569 | 50,798 | 11.0% |
| Agricultural and veterinary sciences | 44 | 3,186 | 1,593 | 10,316 | 15.4% |
| Industrial and information engineering | 59 | 4,865 | 2,433 | 32,086 | 7.6% |
| Total | 67 | 34,776 | 17,383 | 163,518 | 10.6% |

*Table 1: Publications selected according to VTR criteria for 2004-2008 and their representativeness, in each UDA*

The set of publications used for the analysis is identified by taking the best rated publications, in terms of AIR, in numbers equal to half the research staff of each university-UDA. Subsequently, analogous to the criteria of the VTR, each publication is attributed a rating equal to:
- 1 (corresponds to the VTR judgment of "excellent"), if the publication places in the top 20% of the dataset, meaning it has an AIR value greater than 80;
- 0.8 (corresponds to "good"), if the publication places in the AIR range of 60-80;

---

[6] We applied a full counting method: each publication is fully counted for each participating university or UDA.

- 0.6 ("acceptable"), if the publication places in AIR range 40-60;
- 0.2 ("fair"), if the publication has an AIR value below 40.

With these criteria, an average value of rating was calculated for each university-UDA. The descriptive statistics for these ratings are presented in Table 2.

| UDA | No. of universities | No. of "excellent" universities (%) | Rating min | Rating median | Rating st. dev. |
|---|---|---|---|---|---|
| Mathematics and computer science | 57 | 50 (87.7%) | 0.20 | 1 | 0.24 |
| Physics | 59 | 58 (98.3%) | 0.93 | 1 | 0.01 |
| Chemistry | 58 | 56 (96.6%) | 0.90 | 1 | 0.02 |
| Earth sciences | 47 | 35 (74.5%) | 0.64 | 1 | 0.09 |
| Biology | 61 | 54 (88.5%) | 0.20 | 1 | 0.16 |
| Medicine | 53 | 50 (94.3%) | 0.88 | 1 | 0.02 |
| Agricultural and veterinary sciences | 44 | 29 (65.9%) | 0.20 | 1 | 0.24 |
| Industrial and information engineering | 59 | 54 (91.5%) | 0.20 | 1 | 0.16 |

*Table 2: Statistics for university performance ratings in each UDA, based on VQR criteria*

As evidenced in column 3, the results show that the ranking of universities is very flat, because of the high number of joint firsts in the ratings. For all the UDAs, the percentage of "excellent" or top universities is over 50%. Physics represents the most extreme case, where a full 58 universities out of 59 place in first position, with the maximum value of rating. This uniformity towards the top is confirmed by the values for the medians and those for standard deviations, which are near zero. It is clear that results from the analysis are closely linked to the small dimension of the share of products extracted for evaluation: as indicated in Table 1, this is an overall subset of products equal to 12% of the total, with minimums equal to 5.8% for the Physics UDA and 6.6% for the Chemistry UDA, the two UDAs with highest concentration of joint firsts. With such limited shares of total product, the possibility of having publications with impact below the level of "excellent" (the 80$^{th}$ national percentile) is clearly very remote. The results demonstrate failure in planning the first Italian VTR. If all the universities had truly selected their best products for the period 2001-2003, the very limited share considered would not have permitted the evaluation to identify different value among the universities.

**4. University research performance rankings based on VQR criteria**

The upcoming VQR provides that each university will present the two best publications realized by each researcher on staff over the 2004-2008 quinquennium. We simulate this scenario by extracting, from all the 2004-2008 publications present in the ORP, the two publications with maximum AIR as attributed to each faculty member eligible for national assessment. Since the VQR dictates certain exclusions involving young members of research staff and those without a stable faculty role in the five years observed, our analysis likewise excludes researchers with less than 3 years of seniority. Table 3 shows, by UDA, the numerosity of the publications to be evaluated according to these criteria and their representativeness with respect to the total product from the researchers considered.

| UDA | Research staff eligible (a) | Publications to select (b = a x 2) | Total publications (c) | b/c |
|---|---|---|---|---|
| Mathematics and computer science | 3,181 | 6,363 | 13,783 | 46.2% |
| Physics | 2,495 | 4,990 | 22,051 | 22.6% |
| Chemistry | 3,123 | 6,245 | 24,349 | 25.6% |
| Earth sciences | 1,225 | 2,450 | 4,581 | 53.5% |
| Biology | 4,965 | 9,930 | 27,660 | 35.9% |
| Medicine | 10,740 | 21,481 | 50,235 | 42.8% |
| Agricultural and veterinary sciences | 3,066 | 6,132 | 10,222 | 60.0% |
| Industrial and information engineering | 4,682 | 9,364 | 31,748 | 29.5% |
| Total | 33,478 | 66,955 | 161,978 | 41.3% |

*Table 3: Publications selected according to VQR criteria for 2004-2008 and their representativeness, by UDA*

In this case, the set of products subject to evaluation is much more substantial than that of the previous scenario: overall, it is 41.3% of the total, with a maximum of 60% in Agricultural and veterinary sciences and a minimum of 22.6% in Physics. These percentages are aggregate average values by UDA, but encompass considerable variation at the level of the individual researcher. In fact, since the distribution of product by each Italian researcher is very concentrated (29% of researchers produce 71% of the total publications), the percentage of articles to be submitted will be very low for 29% of researchers and very high for the other 71% (Abramo et al., 2011).

Next, analogous to the VQR criteria, every publication is assigned a rating equal to:
- 1 (corresponds to a VQR judgment of "excellent"), if the publication places in the top 20% of the dataset, meaning it has an AIR value greater than 80;
- 0.8 ("good"), if the publication places in the AIR range of 60-80;
- 0.5 ("acceptable"), if the publication places in AIR range 50-60;
- 0 ("fair"), if the publication has an AIR value below 50.

In addition, again in accordance to VQR criteria, a rating of -0.5 is assigned for each publication that is lacking in respect to the number anticipated from the university[7].

Table 4 presents the descriptive statistics for the distribution of average ratings for each university-UDA.

| UDA | N. of Universities | N. of top Universities (%) | Rating min | Rating median | Rating Dev. st. |
|---|---|---|---|---|---|
| Mathematics and computer science | 58 | 1 (1.7%) | 0.10 | 0.62 | 0.15 |
| Physics | 60 | 2 (3.3%) | 0.38 | 0.80 | 0.12 |
| Chemistry | 58 | 4 (6.9%) | 0.52 | 0.83 | 0.10 |
| Earth sciences | 47 | 2 (4.3%) | 0.08 | 0.61 | 0.23 |
| Biology | 62 | 2 (3.2%) | -0.25 | 0.73 | 0.21 |
| Medicine | 54 | 2 (3.7%) | 0.05 | 0.71 | 0.15 |
| Agricultural and veterinary sciences | 48 | 5 (10.4%) | -0.25 | 0.59 | 0.32 |
| Industrial and information engineering | 61 | 5 (8.2%) | -0.25 | 0.69 | 0.22 |

*Table 4: Statistics of university performance ratings based on VQR criteria, by UDA*

As evidenced by column 3, the results show that the classifications of universities, unlike in the first scenario, give a notably lower number of joint firsts in the rankings by area: the highest number (five) is seen in Agricultural and veterinary sciences (10.4%) and in Industrial and information engineering (8.2%). Mathematics and computer science does not show the phenomenon of joint firsts. As evidenced by column 4, a full four UDAs out of nine show negative values for minimum rating, attributed to the presence of universities with a high share of non-productive or low-producing researchers. The median for the university ratings varies from a minimum of 0.59 for Agricultural and veterinary sciences to a maximum of 0.83 for Chemistry. The values of standard deviations vary from a minimum of 0.10 for Chemistry to a maximum of 0.32 for Agricultural and veterinary sciences, and show lower values in correspondence with the higher values of median.

---

[7] In this simulation we assume that a scientist with no publications indexed in WoS has no other research output sto submit.

## 5. Discussion and conclusions

National evaluation exercises provide a comparative measure of performance for a nation's research institutions, and as such represent a stimulus instrument for improvement in productivity, particularly when the results are used to inform selective government funding. In the presence of evaluation criteria that are subject to variation and are not made known with sufficient advance, it could occur that some universities are penalized, rather than rewarded, for following the objectives directed by the criteria known from the preceding exercise.

In this work, to study the phenomenon in detail, we have examined the research performances of universities for the period 2004-2008, object of the next national evaluation, according to the criteria of the first exercise conducted in Italy (VTR), meaning the criteria that are until now known to the universities. We have then compared these performances to those measured under the criteria to be adopted for the upcoming exercise (VQR). The simulations, carried out on data for the hard sciences, show a series of significant discrepancies. And, from the results obtained by applying the old VTR rules, we observe an extreme flattening of the ranking list in each UDA, with a high number of joint firsts at the top of the classifications, which actually impedes any possible distinction among these universities.

Concerning the hypothesis that the bibliometric method could be inadequate for this type of analysis, it should be noted that both evaluation methodologies, bibliometrics and peer review, present advantages and disadvantages. It has been amply demonstrated that there is a positive correlation between the results obtained with peer review and those obtained with bibliometric methods, at least in the fields of the hard sciences (Moed, 2009; Pendlebury, 2009; Abramo et al., 2009; Aksnes and Taxt, 2004; Oppenheim and Norris, 2003; Rinia et al., 1998).

In this case it is legitimate to suppose that these results originate from the methodology applied in the VTR exercise, which rather than recognizing average quality or productivity of an organization, had the objective of identifying and rewarding excellence. But under the criteria of the evaluation, the share of research product evaluated is so limited as to make it impossible to detect significant differences among the universities. If the universities had truly selected their best products for the period 2001-2003, the very limited share included in the evaluation of excellence would not have permitted identification of different values between the universities operating in the various UDAs. However the results obtained under the VQR rules permit better discrimination between universities in each UDA, and the differences between the two scenarios are truly macroscopic. Although it is possible that a part of the polarization of the rankings list in the first scenario could be caused by limits in the bibliometric approach, it is evident that the discontinuity represented by the new criteria of the VQR compared to those of the former VTR implies an upset in the evaluation results.

The Italian situation is certainly an extreme case, however without important consequences in the university system. In fact, the share of funds assigned on the basis of results from evaluation of research activity is very limited. In 2010, this share is 4.9% of total university income. Further, the scheme for awarding these resources actually provides very little reward or punishment, never completely eliminating funding to any university. The situation in nations with competitive higher education systems is much different. For example, the Higher Education Funding Council for England (HEFCE) does allocate no funds to universities placing in the bottom quartile of RAE rankings. Further, the universities with an evaluation profile in the first quartile are assigned (under equal numbers of research staff) funds that is triple that for the universities in the second quartile, and these in turn receive three times that of those in the third quartile. In this scenario, introducing significant discontinuities in the evaluation model will increase the risk of penalizing the universities that, as time passed, have made management choices in keeping with the known evaluation criteria. If Italy were to adopt the same criteria as the HEFCE we would see a situation where, in some areas, a quarter of the universities characterized by the VTR as having a "top" research profile would end up with no funds under the results of the new VQR. At the same time, we would see rewards going to those universities who registered better performance under the new rules that came out at the end of the game. This would disrupt many research organizations, since they would perceive a lack of reward for efforts to pursue policy objectives, as understood from the rules of the game that the policy-maker communicated.

The authors thus call for greater attention from policy-makers to the potential negative consequences of planning national evaluation exercises that first call on organizations to pursue A, but later reward those that achieved B. The impression is that, because of its nature, peer review is not readily compatible with the necessity of setting and communicating assessment model criteria with ample warning prior to the period evaluated. From this point of view, the purely bibliometric approach, although still only applicable to the hard sciences, offers clear advantages of less time for set-up and implementation. In addition, while respecting budget restrictions, it would make it possible to asses all the relevant scientific product from the subjects evaluated. This would, in addition to the evident advantages of precision-robustness and reliability-functionality, eliminate the risk of running into the dilemma that inspired this work.